# Measurement of Wave Electric Fields in Plasmas by Electro-Optic Probe

M. Nishiura, Z. Yoshida, T. Mushiake, Y. Kawazura, R. Osawa[1], K. Fujinami[1], Y. Yano, H. Saitoh, M. Yamasaki, A. Kashyap, N. Takahashi, M. Nakatsuka, A. Fukuyama[2],

*Graduate School of Frontier Sciences, The University of Tokyo, Chiba 277-8561, Japan*
[1]*Seikoh Giken Co. Ltd, Matsudo, Chiba 270-2214, Japan*
[2]*Department of Nuclear Engineering, Kyoto University, Nishikyo-ku, Kyoto 615-8540, Japan*
e-mail address : nishiura@ppl.k.u-tokyo.ac.jp

## Abstract

Electric field measurement in plasmas permits quantitative comparison between the experiment and the simulation in this study. An electro-optic (EO) sensor based on Pockels effect is demonstrated to measure wave electric fields in the laboratory magnetosphere of the RT-1 device with high frequency heating sources. This system gives the merits that electric field measurements can detect electrostatic waves separated clearly from wave magnetic fields, and that the sensor head is separated electrically from strong stray fields in circumference. The electromagnetic waves are excited at the double loop antenna for ion heating in electron cyclotron heated plasmas. In the air, the measured wave electric fields are in good absolute agreement with those predicted by the TASK/WF2 code. In inhomogeneous plasmas, the wave electric fields in the peripheral region are enhanced compared with the simulated electric fields. The potential oscillation of the antenna is one of the possible reason to explain the experimental results qualitatively.

## I. Introduction

Electromagnetic waves in plasma have been studied since early stage as wave physics for plasma heating and diagnostics [1]. In space, the electromagnetic waves have been observed as whistler and Alfvén waves in planetary magnetospheres. These waves propagate in media with complex characteristics. Thermonuclear fusion device based on the dipole field concept was motivated by spacecraft observations in the Jovian magnetosphere [2], and was firstly proposed by Hasegawa [3]. This concept was realized as Ring Trap 1 (RT-1) at The University of Tokyo [4] and Levitated Dipole experiment (LDX) at MIT [5]. In the laboratory magnetosphere, RT-1 uses a klystron with the frequency of 8.2 GHz and a magnetron with the frequency of 2.45 GHz for electron cyclotron heating (ECH) for high electron beta plasma ($\beta_e > 1$) [6, 7]. In RT-1, anisotropic state of ions are studied [8]. For high ion-beta state, ion cyclotron range of frequencies (ICRF) heating was performed in magnetosphere configuration [9]. The antenna excites a slow wave with left-handed polarization in the frequency of a few MHz. The electromagnetic wave propagates along magnetic field lines from high to low field sides, so-called "*magnetic beach*" heating. The ion heating scheme was successfully demonstrated, and resulted in the increase of ion temperature. However the heating efficiency and its wave physics are still unclear.

In the magnetosphere dipole configuration, the curvature of the magnetic field and the plasma production area are different from the conventional linear machines [10-12]. For characterizing electromagnetic waves excited in plasmas, a magnetic probe that loops and coils are made of a metal wire, is used conventionally for detecting the wave magnetic fields [13, 14] in electromagnetic waves and other researches. The merits of this simple method are local measurement, cost effective, and high heat endurance. However, it is difficult to detect the electrostatic components due to potential oscillations by antenna voltage, electrostatic waves, and unexpected mode converted waves.

To avoid the disturbance of the electromagnetic field, a miniature antenna is used. However the antennas still disturb wave fields and the cable connected to the antenna picks up stray wave fields. An electro-optic (EO) sensor for electric field measurement suggests to minimize the disturbance of the wave field where we measure and to reduce the noise mixing from the circumstances. Thus EO sensors have applied to the fields of communications [15], ion thruster [16] and the measure of electromagnetic compatibility (EMC) [17]. In plasma experiments, the EO sensor head that is separated electrically from the detection system is advantageous in measuring excited waves in intense noise fields produced by EC and ICRF heating.

In this paper, we describe the measurement system of wave electric fields based on an EO probe in plasmas, and discuss the experimental results with the help of a finite element method, TASK/WF2 code in the cold plasma theory.

## II. Electro optic probe for the measurement of electric field

There exists two kinds of electric field probes. Both of them are based on Pockels effect in the electric fields; one is the interferometer-type EO-sensor with an optical waveguide containing a miniature antenna on the $LiNiO_3$ substrate [17]. The other is the bulk-crystal-type EO-sensor that detects the change in a polarization degree caused by the change in the refractive index [15] due to the electric fields. The former type of the EO sensor (CS-1403, Seikoh Giken Co. Ltd) is used in this study. Figure 1 shows the schematic of inside the sensor head made of the $LiNiO_3$ crystal substrate. The single-mode optical-fiber is spliced to the optical waveguide fabricated on the substrate. The sensor head is covered by acrylic resin case for shock protection. The size of the cuboid is 6×5.5×23.5 mm. The electric field and frequency is covered from 1 V/m–25 kV/m and 100 kHz–10 GHz, respectively.

For the electric field measurement in RT-1, the sensor head with the optical fiber is placed inside a quartz tube with the thickness of 5 mm, as is shown in Fig. 2. The quartz tube is connected to a stainless steel tube at the location of 24 mm from the quartz tip by a Viton O-ring for the vacuum seal. As the temperature of the EO sensor is limited below 60 ˚C, foregoing to the electric field measurement, we make sure that the temperature does not exceed 60 ˚C at the same plasma discharge.

The measurement setup of electric field is shown in Fig. 3. Polarization maintaining (PM) and single mode (SM) fiber cables connect the components. Polarized laser light (IDPHOTONICS, the wavelength of 1.6mm and the power of 16 dBm) is delivered to the EO sensor (Pockels crystal is mounted) via optical fibers and a circulator. The laser light is divided into two photo waveguides fabricated on the $LiNbO_3$ crystal substrate; the photo waveguide system forms the interferometer with amplitude modulation of the optical signal. One gives rise to the phase delay due to the change in the refractive index of the $LiNbO_3$ by applied electric field. It is picked up by a printed dipole antenna on the $LiNbO_3$. The other one passes through the photo waveguide without a phase delay. Both of laser lights are reflected back at the end mirror in the EO sensor, and are merged into one beam at the SM fiber. The InGaS PIN photo detector (Newport, wavelength 1000–1650 nm, bandwidth 12.5 GHz) detects the interfered signal through the circulator. The output signal of the photo detector is monitored by a spectrum analyzer (Anritsu, MS2720T, bandwidth 9 kHz–20 GHz) determining the time resolution of a hundred milliseconds. The EO sensor head mounted on RT-1 is separated electrically by a SM fiber with 20 m in length. The rest of the components is placed at the control room for the reduction from electromagnetic noises.

The EO sensor system is calibrated to obtain the absolute intensity of electric field in this frequency range. Figure 4 shows the radiation pattern of the EO sensor which was measured in the shield box surrounded by electromagnetic absorbers. The output waves of a synthesizer was introduced into the shield box. The EO sensor used here was directional to the transverse to the optical fiber axis. The EO sensor head was rotated on the optical fiber axis to obtain the radiation patterns and calibration factors for the frequencies of 1, 2, and 3 MHz. The directional sensitivity of 0 and 180 degrees is about one order of magnitude higher than that of 90 and 270 degrees. The calibration factors are a few dB fluctuation up to 10 GHz.

## III. Measurement of wave electric fields for ion heating in magnetosphere plasmas

Figure 5 shows the top and cross sectional views of the magnetosphere plasma device RT-1. The double loop antenna for ICRF heating is mounted on the center stack with supporting rods which are insulated electrically from the vacuum vessel. For magnetic beach heating, the lower and upper loop antennas are located at $\omega/\Omega_{He^{2+}} \sim 0.58$ and $\omega/\Omega_{H^{+}} \sim 0.66$ in the inner high field side, respectively. The schematic of the antenna with the current directions are indicated in Fig. 6. One end of the antenna is connected to the current feedthrough for rf power fed from the rf power supply of 10 kW nominal output. The matching box is placed in between the feedthrough and the rf power supply.

The EO sensor is implemented to measure the electromagnetic waves in RT-1 plasmas. The supporting rod enclosing the EO sensor is inserted from the top port #5-T-0 at the radial position R = 0.245 m. The sensor head travels to obtain the vertical profile of wave electric field in the theta direction $E_\theta$. The EO sensor can detect the stray radiations of electromagnetic waves from the EC and ICRF heating in the RT-1 vacuum vessel. Figures 7 (a) and (b) show the typical spectra that come from MHz and GHz ranges during plasma shots. These spectra are acquired by the spectrum analyzer with the time resolution of a few hundred milliseconds. The intensity of the wave electric field is obtained by subtracting the noise floor from the peak, and by multiplying the subtracted value by the calibration factor that is measured in Fig. 4.

To validate the EO sensor system, the measured electric fields in air and plasma are compared with those calculated by TASK/WF2 code which is based on a finite element method to solve Maxwell's equations in a cold plasma model. The vacuum vessel of RT-1, the levitation coil, and center stack were set to a conductor boundary in two dimensional coordinate. Quantitative comparisons require the antenna current which was measured by a current transformer at the current feedthrough close to the antenna. In the air, as plotted in Fig. 8, the profile of $E_\theta$ was measured at R = 0.245 m. At Z = 0.22 the EO sensor is contact close to the double loop antenna. The measured $E_\theta$ was in good quantitative agreement with the simulation. In this case, the antenna was excited at the frequency of 3 MHz. In the simulation, $E_\theta$ is calculated at the antenna current of 1 A, and is multiplied by the measured one to obtain the absolute value of $E_\theta$. The measured antenna current of 5.0 A was used for the simulation. The detected signal was almost disappeared far from $Z \geq 0.32$ m due to the noises.

The EC heating started up the discharge, and at the onset of the discharge the double loop antenna excited electromagnetic waves propagating in plasmas as a slow L wave. The wave propagates parallel to the magnetic field from high to low magnetic field sides, and is absorbed at the ion cyclotron layer in lower magnetic fields. The ICRF heating started, while the EC heating plasma retains 1 second discharge. The EO sensor position was moved vertically shot by shot at R = 0.72 m (5-T-2 ports). Figure 9 shows the EO sensor, the double loop antenna, and the levitation superconducting coil inside the vacuum vessel of RT-1.

The EC heating with the power of $P_{ECH}$ = 12.2 kW started up the discharge at the helium gas pressure of 2.1 mPa. The ion temperature of $He^+$ was monitored by the Doppler broadening of bulk helium ions (He II, $\lambda$ = 468.57 nm) and impurity ions (C III, $\lambda$ = 464.74 nm) on the equatorial plane [8, 9]. In the discharge, the ion temperatures were $T_i$(HeII) ~ 10eV and $T_i$(CIII) ~ 30eV with spatial profiles. The line averaged electron densities, $n_0$ for the central and $n_{edge}$ for the peripheral chords were measured by 74 GHz interferometers IF1 with the horizontal line of sight at R = 0.45 m, and IF3 with the vertical line of sight at R = 0.7 m, respectively. The electron densities were $n_0 = 1\times10^{17}$ $m^{-3}$ and $n_{edge} = 1\times10^{16}$ $m^{-3}$. The discharge conditions appeared in later figures are also summarized in the table 1. In the TASK/WF2 simulation, the experimental parameters were entered to calculate $E_\theta$. We assumed that the electron density profile is dominated by a flux surface function of 1/R in the dipole field of RT-1 with the measured values for $n_0$ and $n_{edge}$. Using the above measured quantities, the power absorption of excited waves and induced electric field in plasmas are calculated by TASK/WF2 code. From the result in Fig. 10, the power-absorption area for $He^{2+}$ exists between the upper and the lower loops. Hence the excited slow L wave at the lower loop is only absorbed at the ion cyclotron resonance layer for $He^{2+}$, and is not for $He^+$.

From Fig. 11, we found that the measured $E_\theta$ is higher than the simulated one at $Z \geq$ 0.35 m where the last closed flux surface is located. In contrast, it decreased and was the same level at Z = 0.3 m. From the above results, it is found that the simulated $E_\theta$ is lower at the outside and comparable at the inside of plasmas. A possible cause should be considered. The effect on the antenna potential was ignored in the TASK simulation. The double loop antenna was oscillated with a voltage of a few kV at an rf power input of 10 kW. Therefore the electrostatic oscillation might provoke the discrepancy in the simulation at the outside of the confinement region.

Apart from the above reason for the discrepancy of field strength, the electron density in front of the antenna might not be a practical one. To verify the edge density profile, the edge electron density was measured by the double probe at the same port R = 0.72 m. The edge electron density $n_{edge}$ by the electric probe with the sampling speed of 100 kHz was the same order of magnitude as that by the interferometer in this operation. The edge electron density was smoothly connected to the core plasma across the last closed flux surface at Z = 0.35 m. The measured profile of $E_\theta$ in Fig. 11 cannot be explained only from the profile of $n_{edge}$. The other possibilities are discussed in the next section.

The wave electric fields were measured in some conditions listed in the table 1. Figure 12 shows the profiles of wave electric fields in helium plasmas with the fill gas pressures of 2.1 mPa and 4.4 mPa. The input powers for EC heating and ICRF heating were almost the same in two cases. In the case of 4.4 mPa, the electron density from the core to the edge is higher than those in the case of 2.1 mPa. Higher electron density becomes lower $E_\theta$.

The wave electric fields along the line of R = 0.72 m were measured by applying the ICRF frequency of 2 MHz in hydrogen and in helium plasmas in Fig. 13. To measure the ion temperature small amount of helium was mixture in hydrogen plasmas. The electric field strengths at Z = 0.35 m in both cases were higher than that in 3 MHz case, and were 185 V/m for hydrogen and 95 V/m for helium plasmas, respectively. This is caused by the coupling between the antenna and the plasmas, because the antenna current was 280–296 A for 2 MHz that is 1.2 times for 3 MHz. The profiles of electric fields had the similar shape outside the separatrix, as discussed before. In the hydrogen plasma, the slow L wave excited by the upper loop antenna experiences the power-absorption layer at the measured positions. We still need the careful study on the relation between the observed electric field and the ion heating.

## IV. Electric field and potential profiles excited in plasmas

The $E_\theta$ measured along the Z direction forms the local maximum near the last closed flux surface. The present model in TASK/WF2 code cannot explain the experimental profile of $E_\theta$, as shown in Fig. 11. We consider the reasons to interpret the measured $E_\theta$. The incomplete separation of $E_\theta$ from other components might arise in the signal of the EO probe, although $E_r$ and $E_z$ are one order of magnitude lower and insensitive to the $\theta$ direction. This case requires the strong $E_r$ and $E_z$ to explain the measured $E_\theta$.

The excitation of electromagnetic waves has been modeled by feeding the external current density $J_{ext}$ that the antenna produces, and Maxwell's equations are solved for the wave propagation in plasmas. However since the intense electric field is induced by the sinusoidal antenna potential associated with charged particle fluctuations, the TASK/WF2 code is modified to simulate the situation. The wave electric field in RT-1 is calculated separately as $J_{ext}$ with the electric charge density $\rho_{ext} = 0$, and $J_{ext} = 0$ with $\rho_{ext}(\neq 0)$ attracted due to the electrostatic potential of the antenna. The current density $J$ and the electric charge density $\rho$ derived from internal plasmas can be expressed as

$$A \cdot \boldsymbol{x} = \boldsymbol{b}$$
$$= \begin{pmatrix} J \\ \rho \end{pmatrix} + \begin{pmatrix} J_{ext} \\ \rho_{ext} \end{pmatrix} ,$$

where the matrix $A$ represents the dielectric constant and the permeability in media, and $\boldsymbol{x}$ plasma parameters related to densities and temperatures.

The electric fields inside the vacuum vessel of RT-1 are calculated for $J_{ext}$ with $\rho_{ext} = 0$. From the results in Fig. 14, the double loop antenna radiates the strong $E_\theta$ for slow wave excitation. In addition, the antenna produces the strong $E_r$ and $E_z$ locally in the region between the levitation coil and the center post as well as in the peripheral region. Since the localized electric field perpendicular to the magnetic field increases the rate of electron heating, this result suggests that Faraday shield is implemented to avoid the localized electric fields.

Figure 15 shows that the synthetic profile of $E_\theta$ evaluates the electric field detected by the EO probe. The profile of $E$ is relatively discussed, for example, when we assume that $E_r$ and $E_z$ of 10% interfere into $E_\theta$ due to the incomplete separation. Although the sum of $E_\theta$ and $0.1(E_r + E_z)$ increases the entire signal level, no steep decrease in the electric field cannot be appeared at Z < 0.35.

The potential oscillation of the antenna is simulated to explain the additional electric field in the peripheral region. Based on the condition that $J_{ext} = 0$ with $\rho_{ext}(\neq 0)$, two dimensional profiles of electric fields and the potential are calculated in Figs. 15 and 16, respectively. The difference in the electric field between Fig. 14 and Fig. 16 appears in $E_\theta$ near the double loop antenna. In addition, the electric charge on the antenna surface induces the strong electric field behind the double loop antenna where ion heating is not necessary. This fact predicts the direction of improved antenna; a Faraday shield structure would be designed to avoid a strong field, and thus the efficient slow wave heating of ions would be expected. The absolute value of $E$ is scaled to the effective antenna voltage of 2.1 kV at the antenna position. The obtained scale factor becomes the electric field too high to explain the measured $E_\theta$. The synthesized profile of $E$ detected by the EO probe also forms similar to that in Fig. 15.

## V. Summary

The EO sensor system was demonstrated to measure the electric fields and analyze the ICRF heating in the laboratory magnetosphere plasmas. The excited wave electric fields were detected in the range of 3–200 V/m in plasmas far from the double loop antenna. The simulation predicts accurately the measured electric fields in the air. Meanwhile, the observed $E_\theta$ in plasmas, particularly outside the last closed flux surface, has the discrepancy between the measurement and the simulation based on the cold plasma theory. An antenna-potential oscillation associated with the antenna excitation may suggest the observed discrepancy in our experiments. The improved structure of the antenna would enhance the efficiency of ICRF heating.

**Acknowledgment**

This work is performed with the support and under the auspices of the NIFS Collaboration Research program (NIFS15KOAH034) and JSPS KAKENHI Grant No 23224014.

Table 1. Discharge conditions and plasma parameters for Figs. 8 and 9.

| Gas [mPa] | $P_{EC}$ [kW] | $P_{ICRF}$ [kW] | $f_{ICRF}$ [MHz] | $I_{ant}$ [A] | IF1 [$m^{-3}$] | IF2 [$m^{-3}$] | IF3 [$m^{-3}$] | Remarks |
|---|---|---|---|---|---|---|---|---|
| He, 2.1 | 12.2 | 6.8 | 3 | 252 | $1.6\times10^{17}$ | $3.0\times10^{16}$ | $1.0\times10^{16}$ | Red circles in Figs. 10 and 11 |
| He, 4.4 | 12.4 | 7.0 | 3 | 250 | $3.9\times10^{17}$ | $7.3\times10^{16}$ | $1.5\times10^{16}$ | Blue circles in Fig. 11 |
| He, 4.6 | 13 | 9.2 | 2 | 296 | $3.1\times10^{17}$ | $5.2\times10^{16}$ | $2.4\times10^{16}$ | Blue circles in Fig. 12 |
| H 8.0 He0.6 | 13 | 10 | 2 | 280 | $1.8\times10^{17}$ | $4.2\times10^{16}$ | $1.8\times10^{16}$ | Red circles in Fig. 12 |

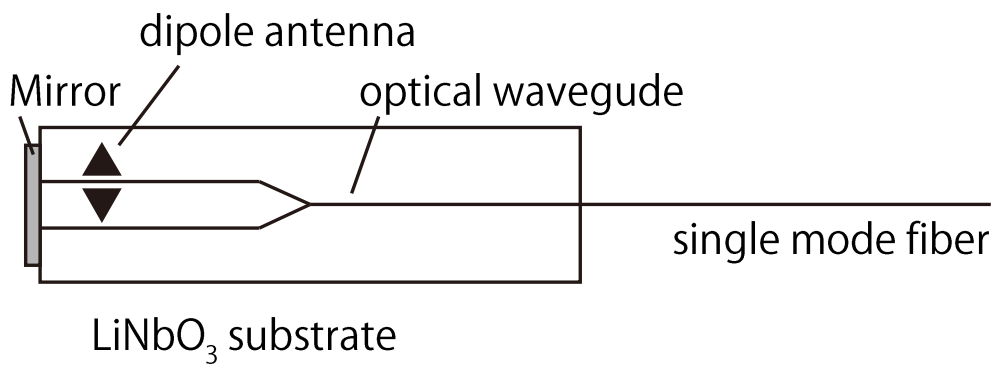

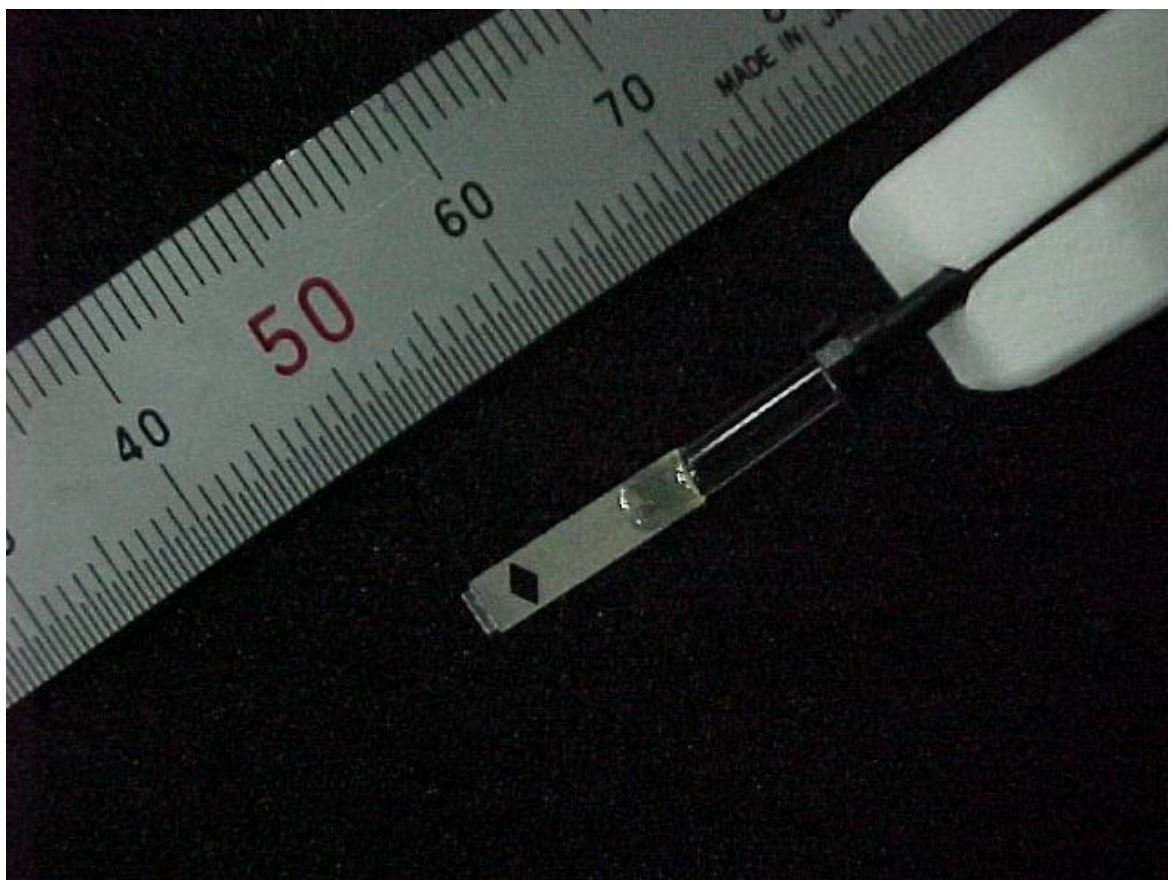

Fig. 1 The inside structure of the EO sensor tip. The tip in the right picture is normally protected by the acrylic resin case for shock.

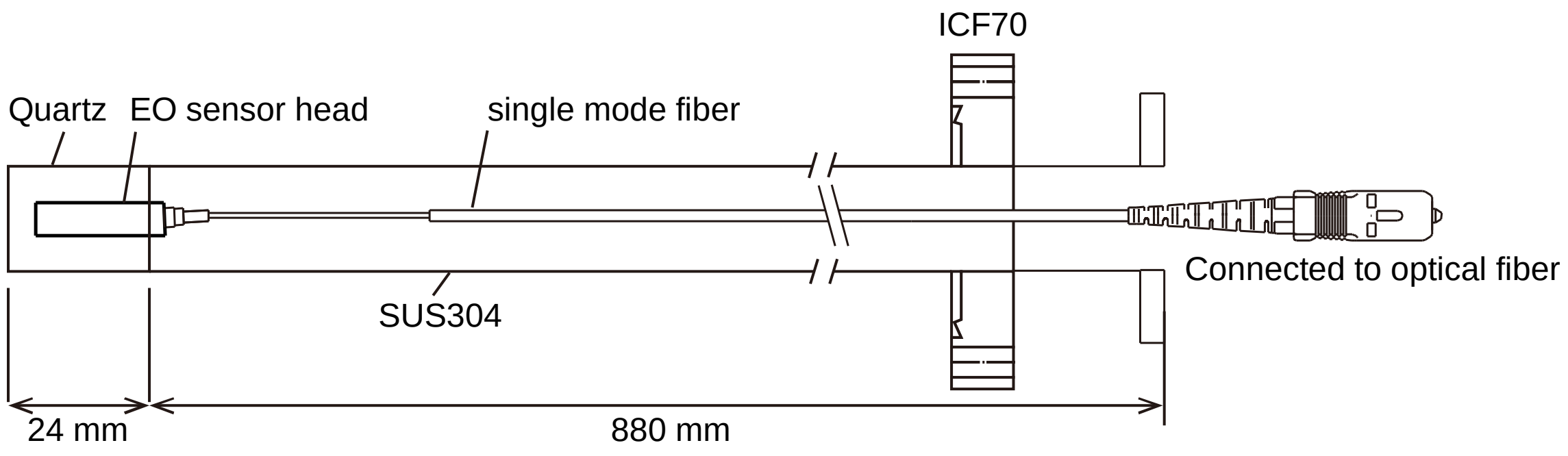

Fig. 2 Schematic of an EO sensor mounted in a shaft to insert into RT-1 plasmas. The shaft is mounted on a motor drive system at the ICF70 flange to measure the spatial profile.

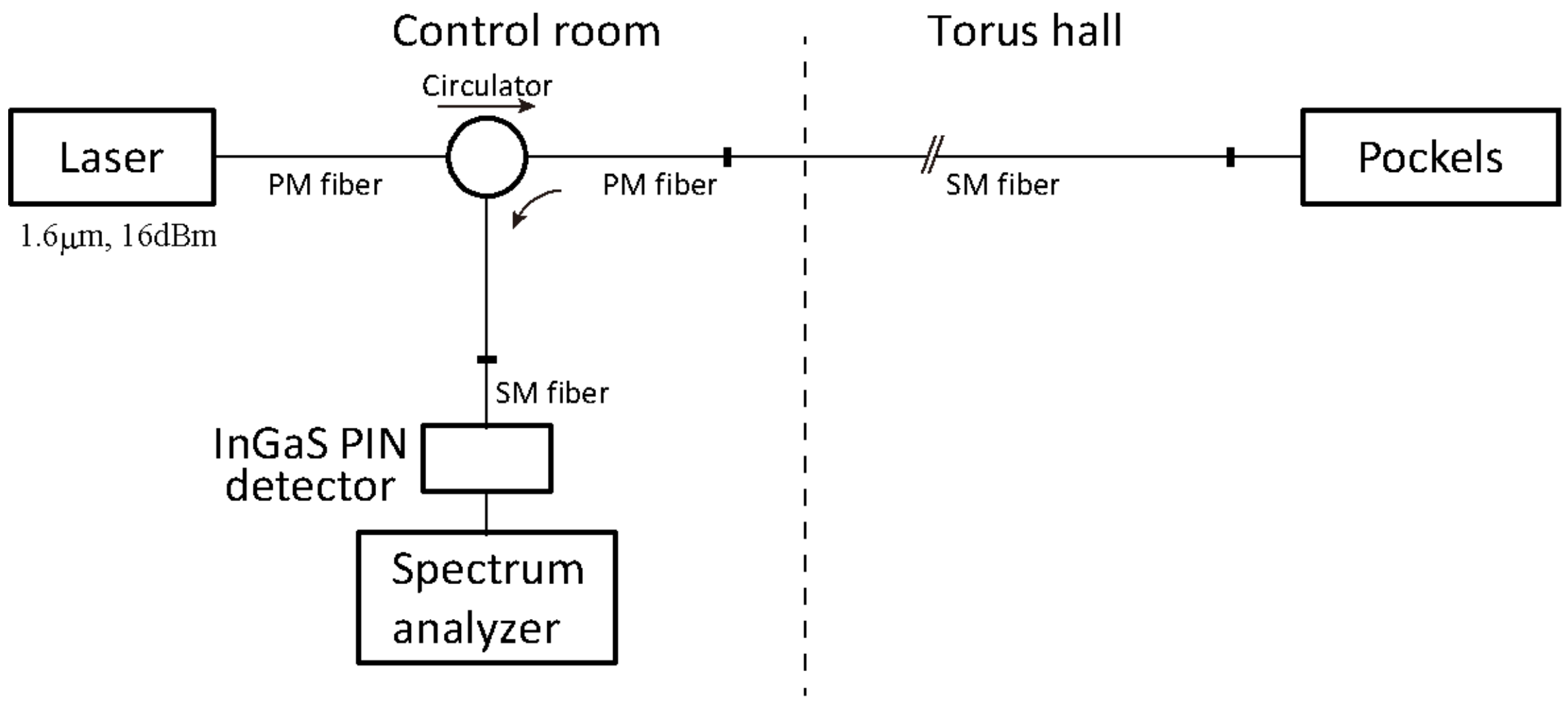

Fig. 3 Measurement setup of electric field with the EO sensor.

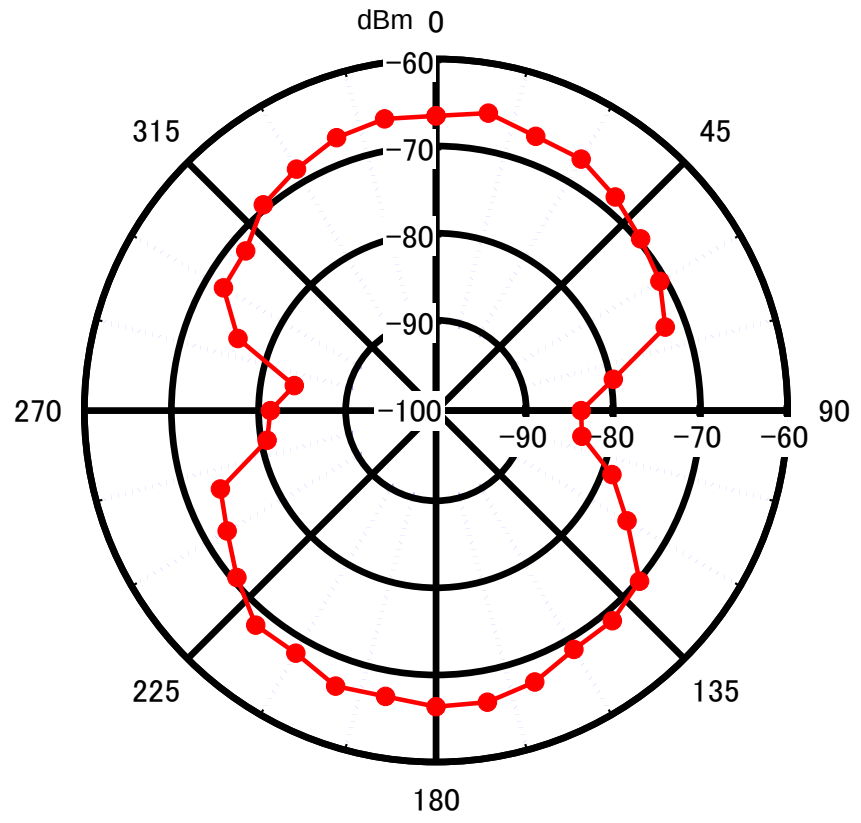

Fig. 4 Radiation pattern of the EO sensor for the frequency of 3 MHz.

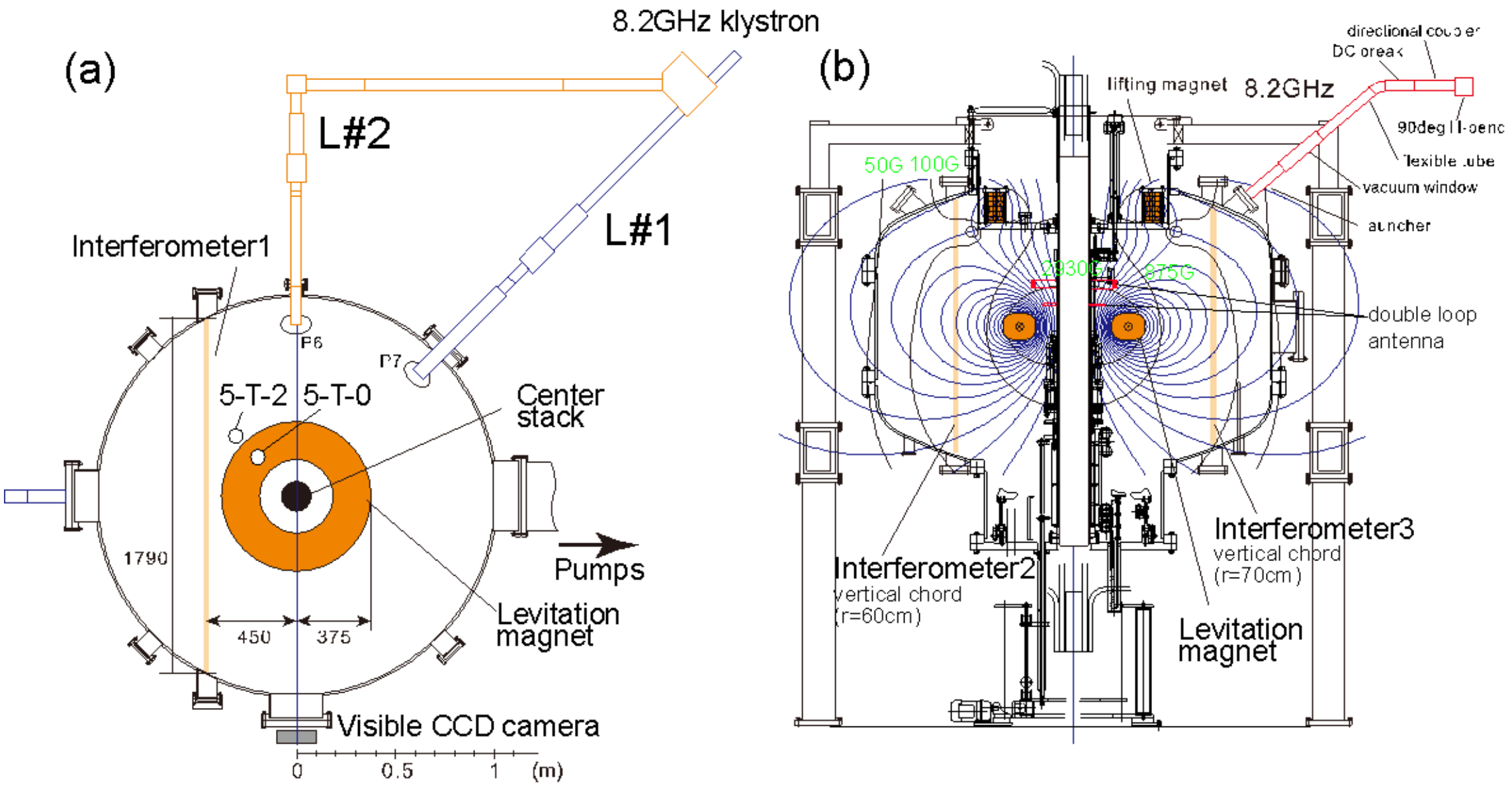

Fig. 5 Magnetosphere plasma device RT-1. (a) top view and (b) cross sectional view. For the electric field measurement, the EO sensor head was inserted from 5-T-0 (R = 0.245 m) and 5-T-2 ports (R = 0.72 m). The double loop antenna for ICRF heating is mounted on the center stack.

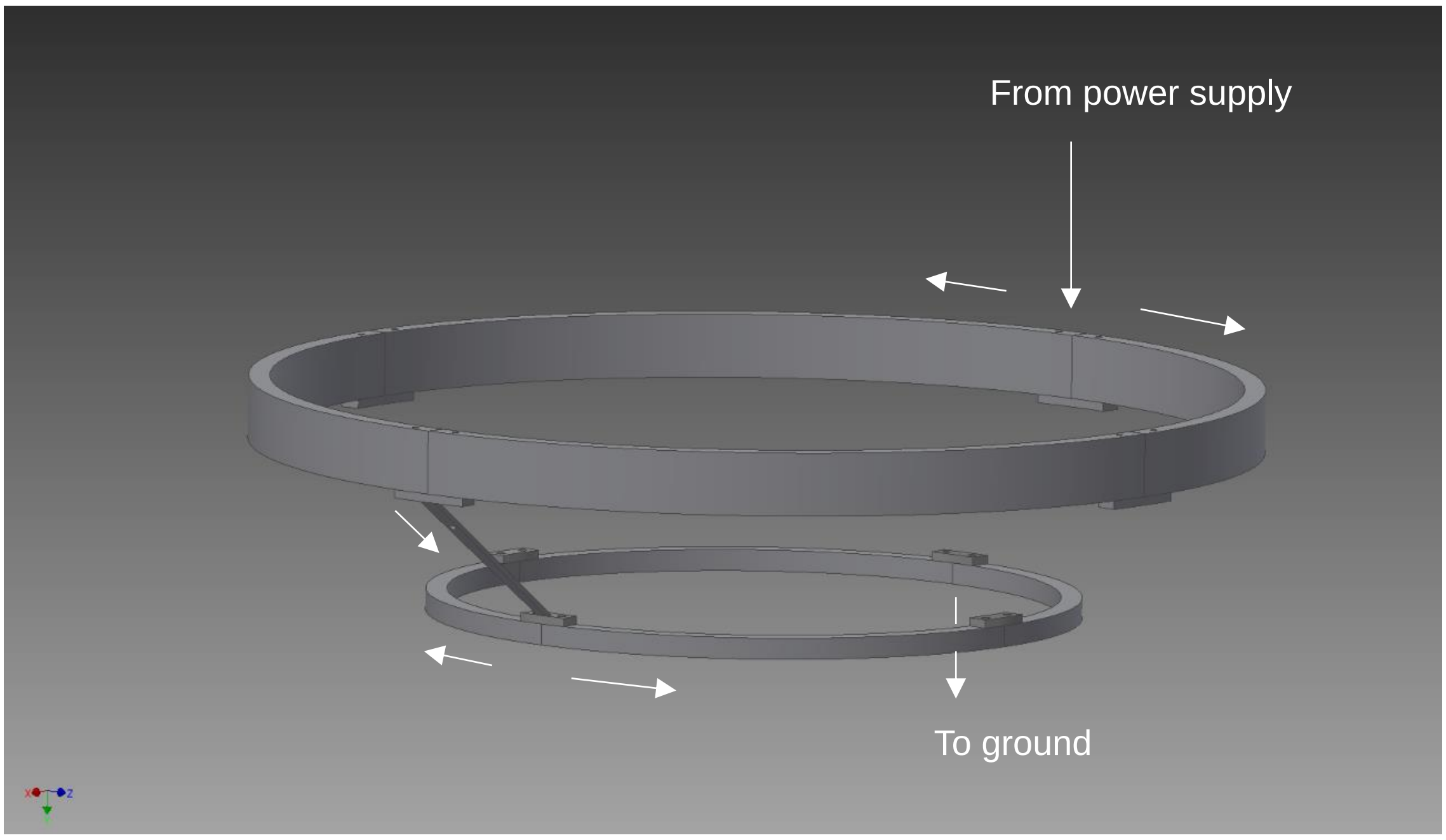

Fig. 6 Double loop antenna and rf current flow for ICRF heating in RT-1.

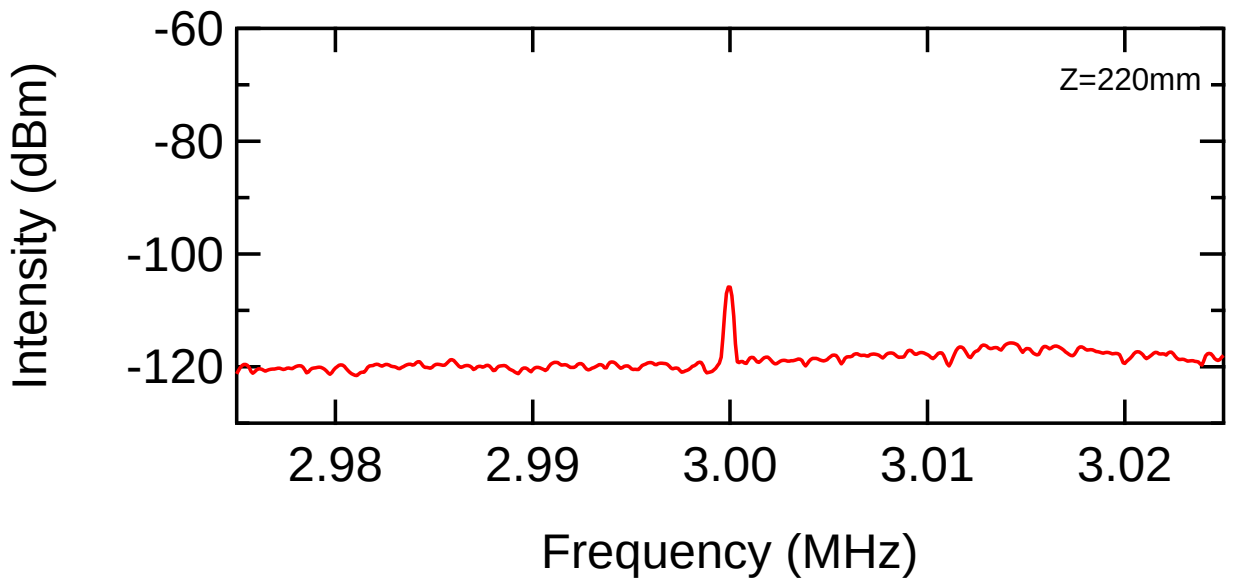

(a) Spectrum of ICRF wave measured by the EO sensor.

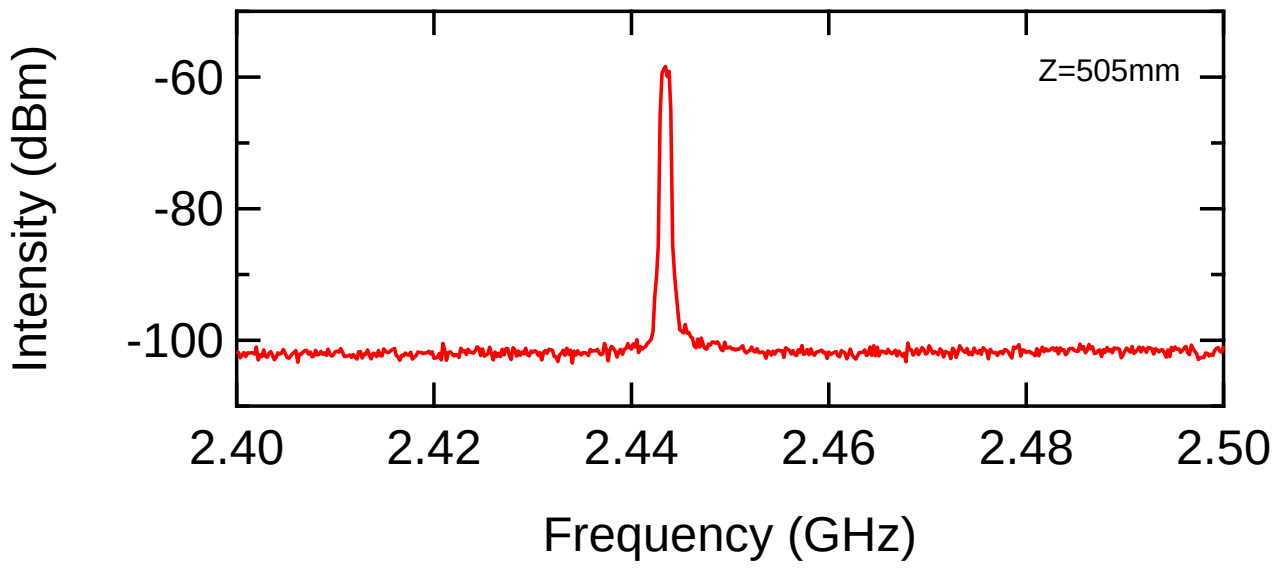

(b) Spectrum of EC wave measured by the EO sensor.

Fig. 7 Typical spectra in (a) MHz and (b) GHz ranges during EC heating discharge.

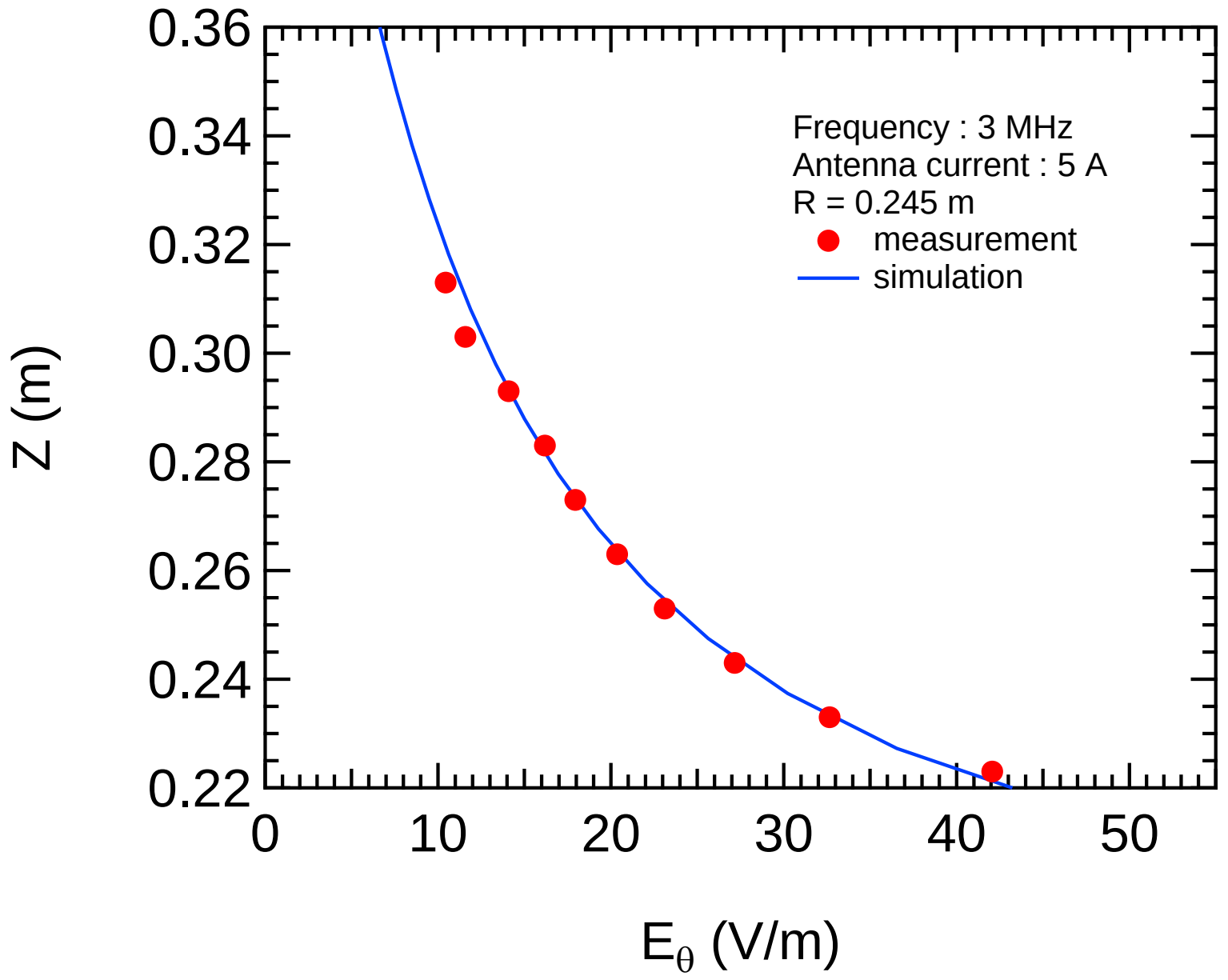

Fig. 8 Electric field $E_\theta$ for 3 MHz measured by the EO sensor (closed circles) in RT-1 in the air.

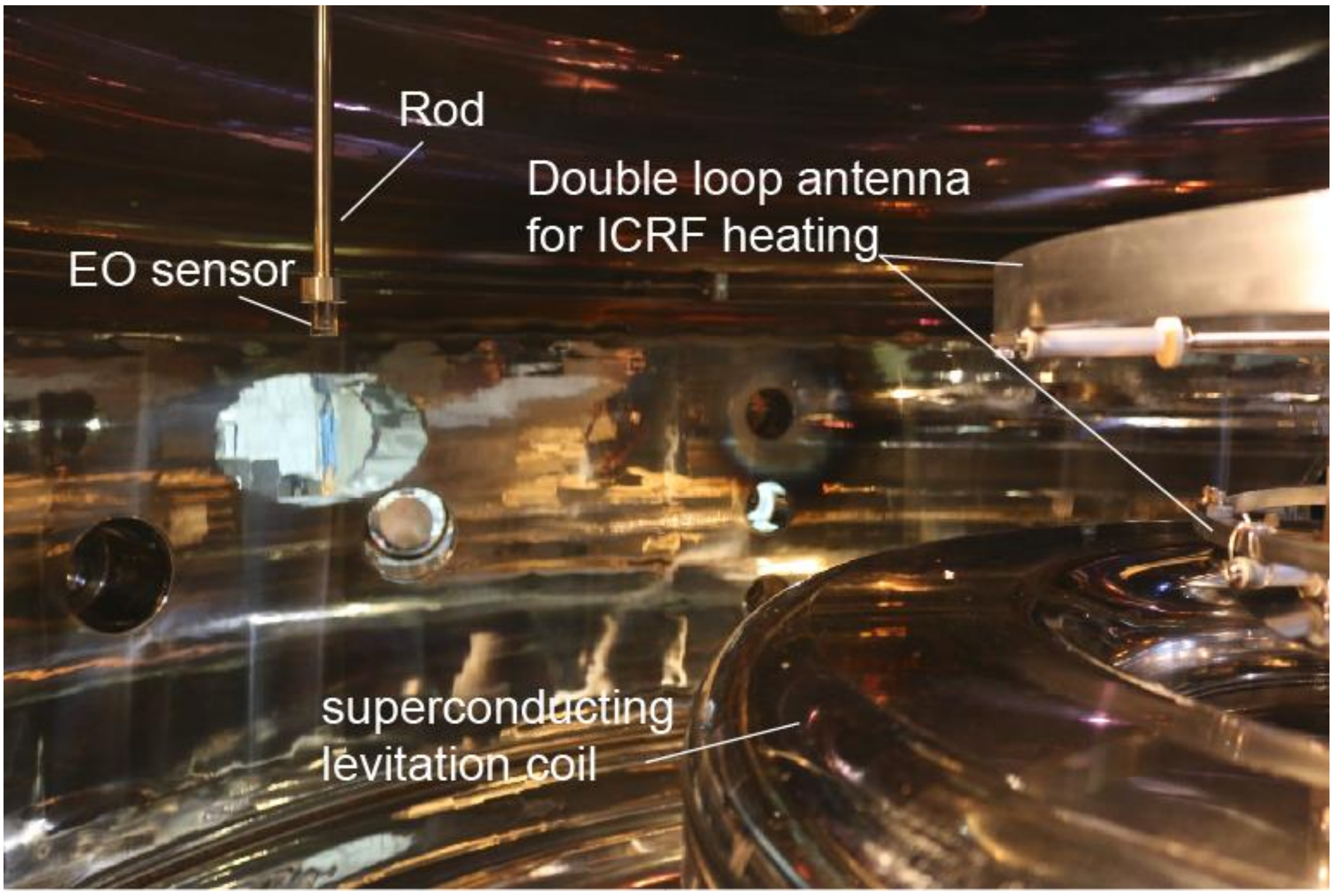

Fig. 9 EO sensor mounted inside the supporting rod is inserted from the top port. In vessel of the RT-1 the levitation superconducting magnet and the double loop antenna for ICRF heating are located.

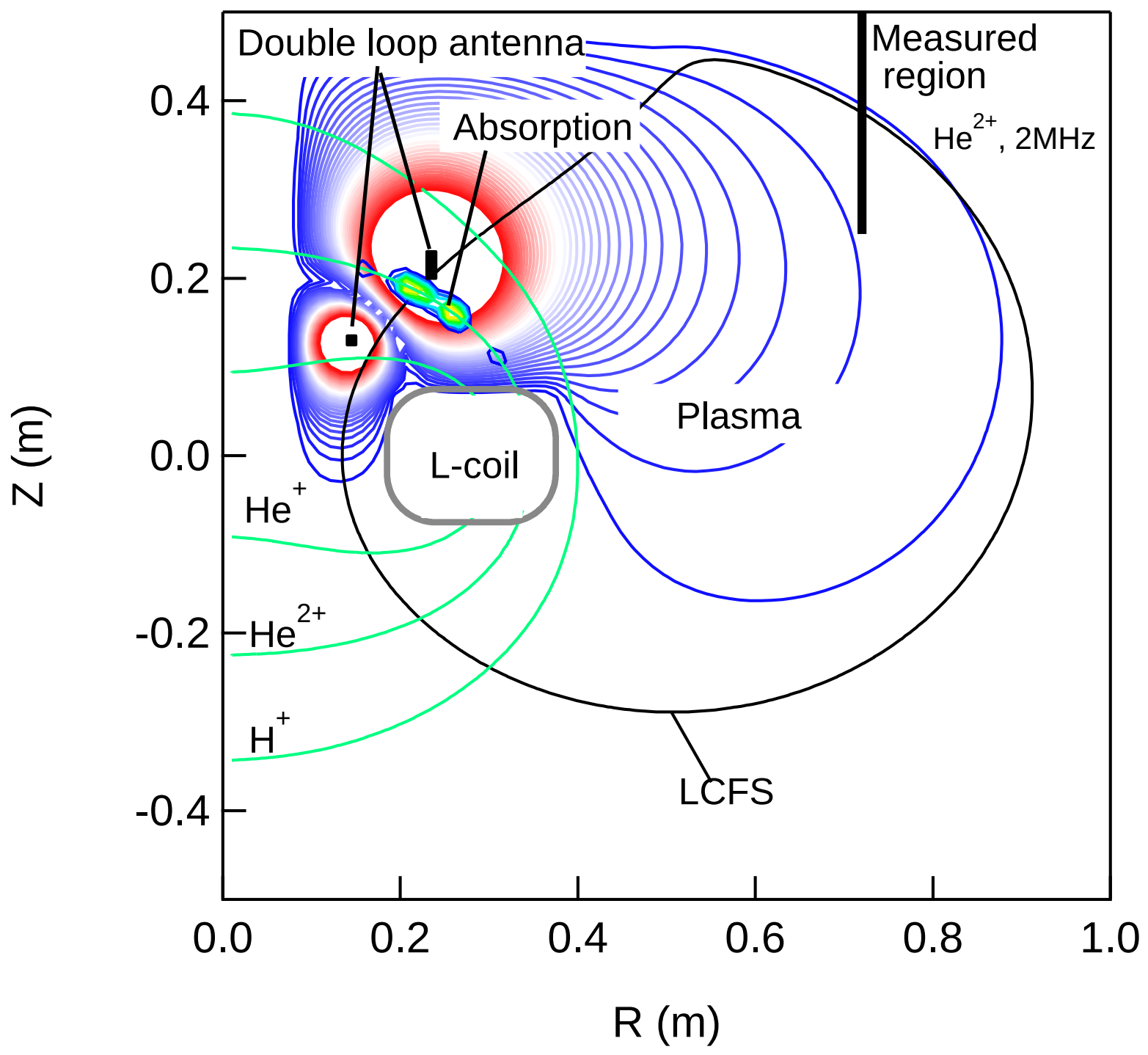

Fig. 10 The contour of electric field $E_\theta$ excited by the double loop antenna in helium plasma. The power absorption area for $He^{2+}$ is also plotted as contour. The ion cyclotron layers for $H^+$, $He^{2+}$, and $He^+$ are depicted.

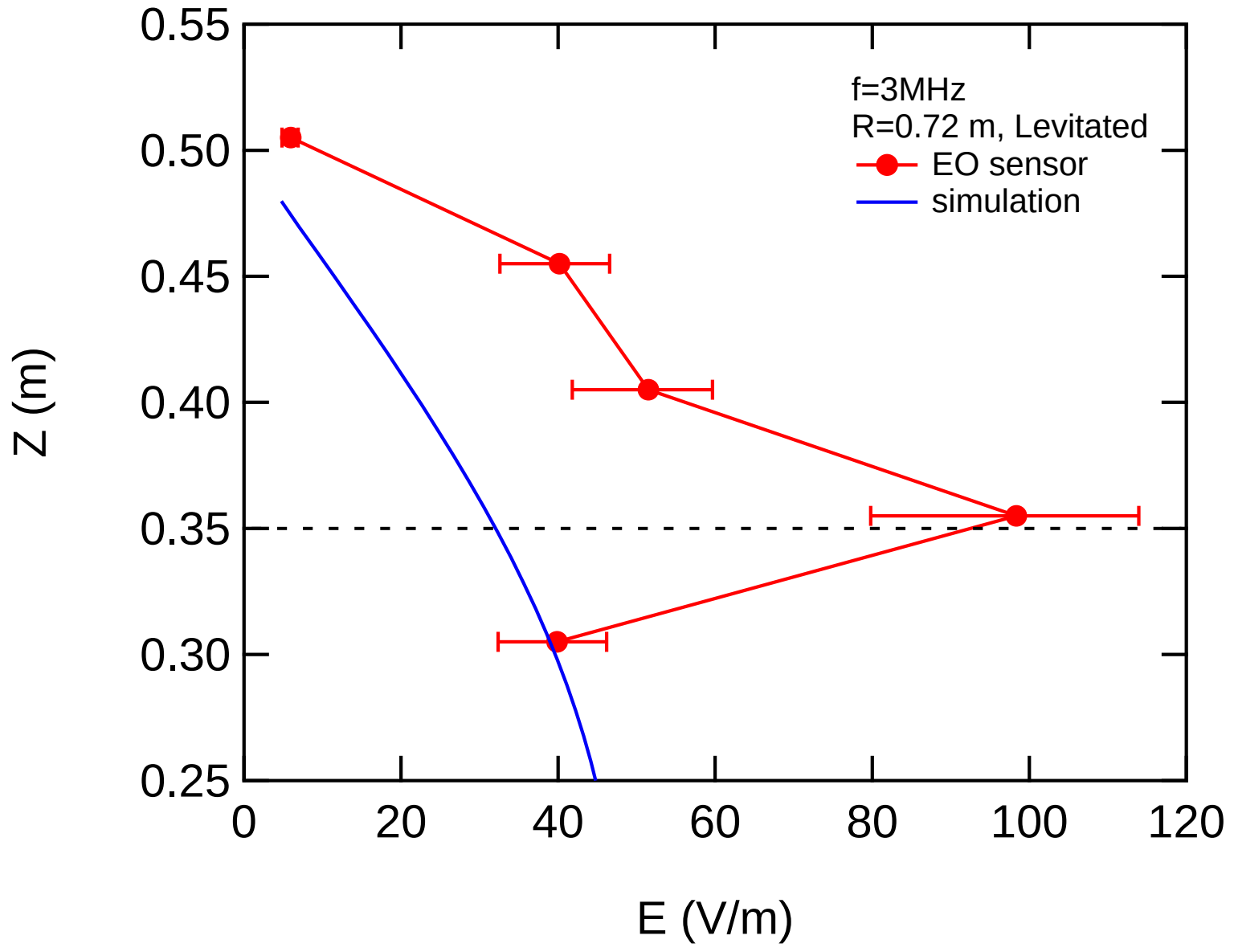

Fig. 11 The EO sensor measured $E_\theta$ (closed circles) along R = 0.72 m vertically in helium plasma. The ICRF power of 7 kW (in right) is applied. The separatrix is located at (R, Z) = (0.72 m, 0.35 m) (broken line). The measured antenna current of 252 A was used as the input parameter for the simulation.

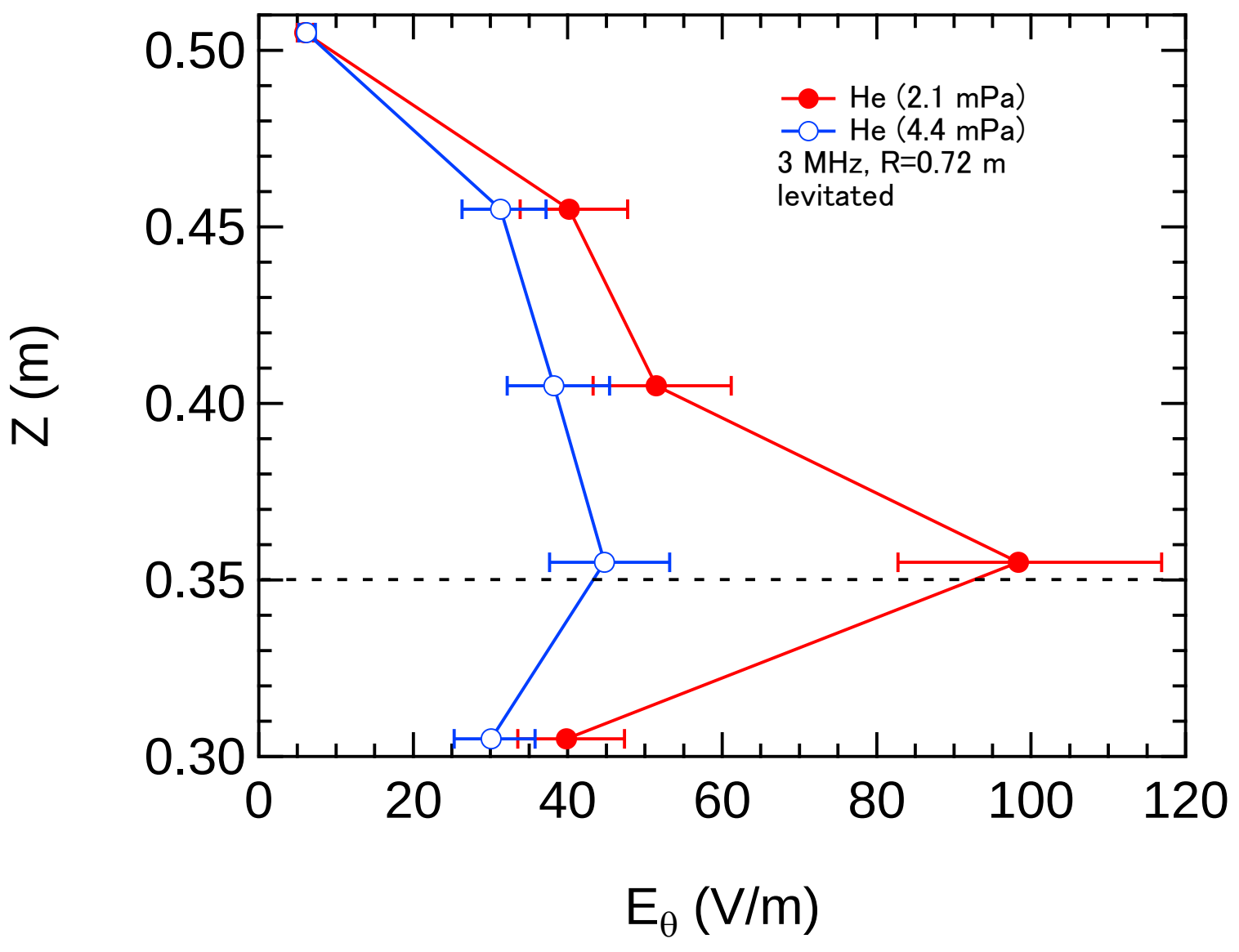

Fig. 12 Electric field $E_\theta$ measured by the EO sensor in helium plasmas. The helium gas pressures were 2.1 mPa (closed circle) and 4.4 mPa (open circle).

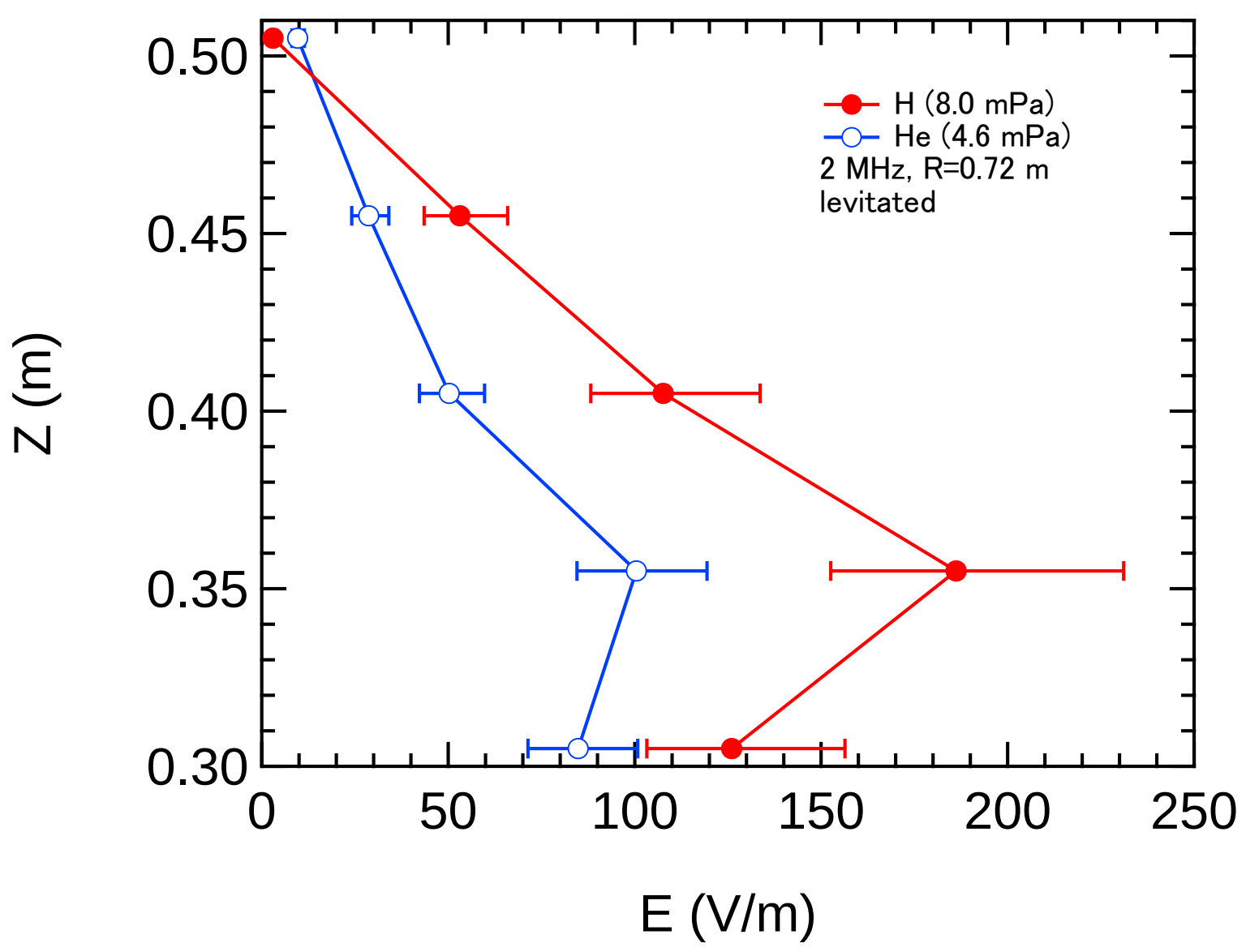

Fig. 13 Electric field $E_\theta$ measured by the EO sensor in plasmas. The fill gas pressures were 8.0 mPa (closed circle) for hydrogen and 4.6 mPa (open circle) for helium, respectively.

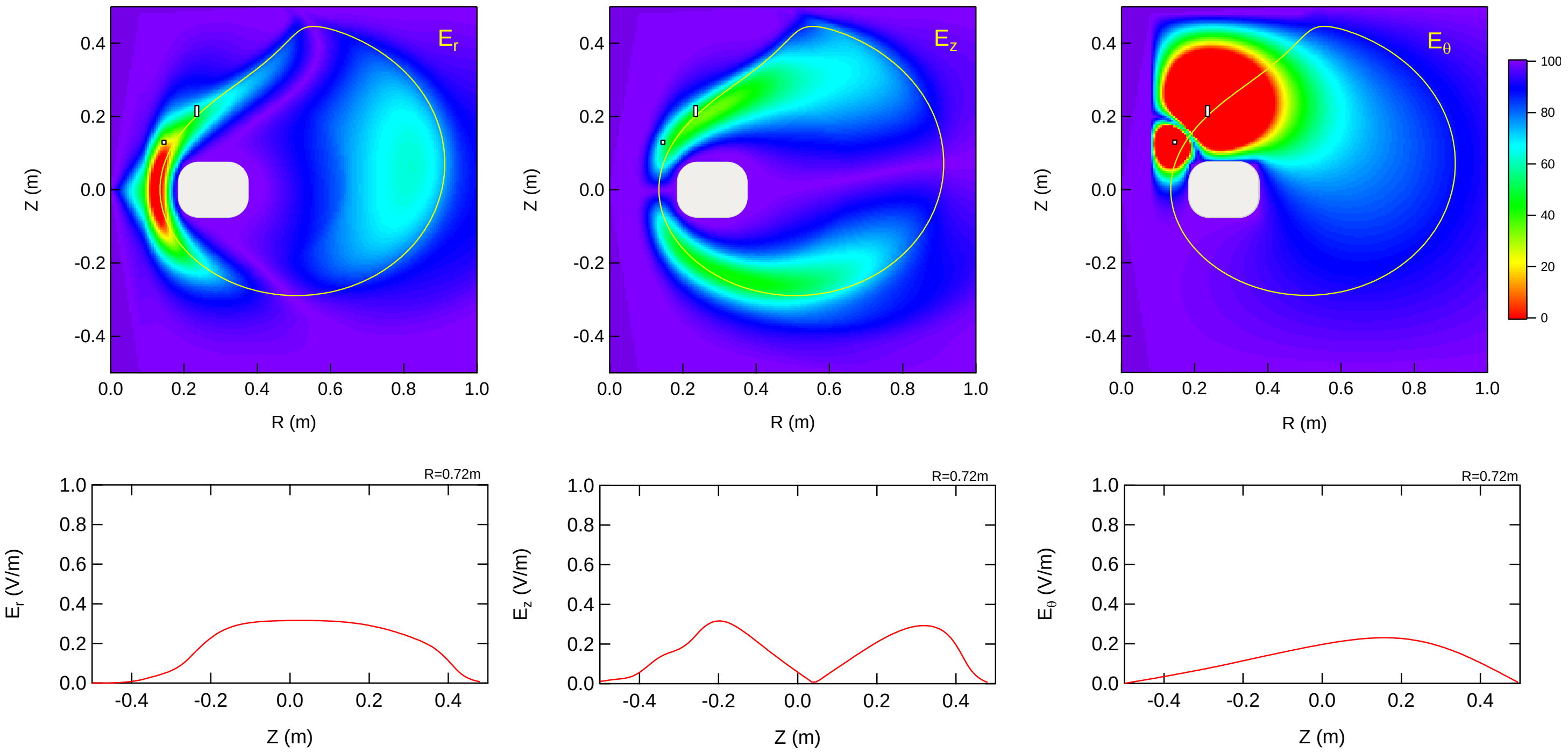

Fig. 14 Two dimensional profile of the electric field in RT-1 excited by the double loop antenna. The current density for upper and lower loops are set to $J_{ext}$ = +252 A and -252A, respectively. The profiles of $E_r$, $E_z$, and $E_\theta$ along the line of R = 0.72 m in RT-1 are plotted from the 2D profiles.

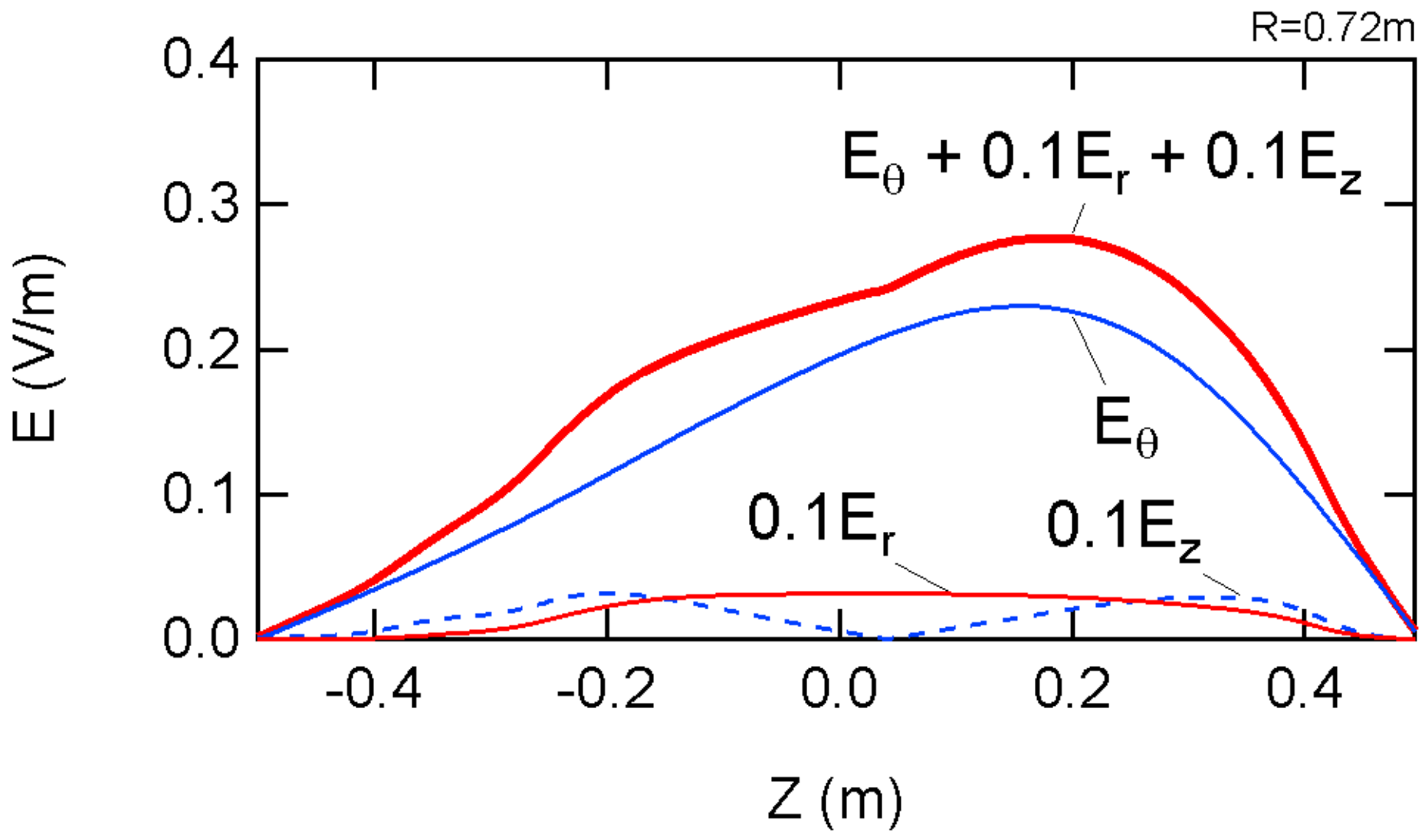

Fig. 15 The profile of the electric field is synthesized along the line of R = 0.72 m, if $E_r$ and $E_z$ of 10% interfere into $E_\theta$ due to the incomplete separation. The wave fields excited by the double loop antenna with the current densities $J_1$ = 1A for the upper loop and $J_2$ = -1A for the lower loop. The result is extracted from the profile of the electric field on the (R, Z) plane in Fig. 14.

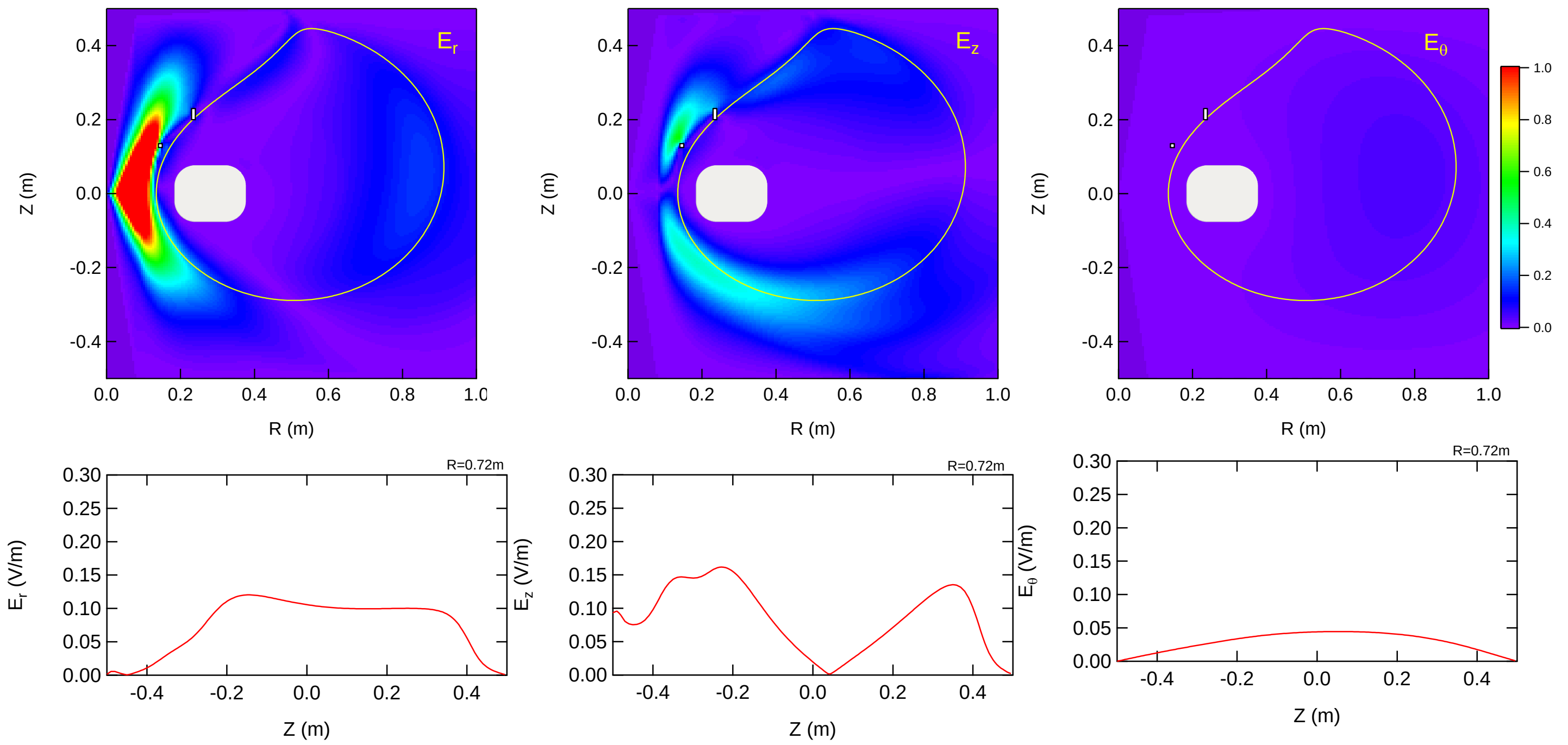

Fig. 16 Two dimensional profiles of the electric field in RT-1 excited by electric charges on the double loop antenna. The antenna currents for upper and lower loops are set to $J_{ext}$ = 0 A. The profiles of $E_\theta$, $E_r$, and $E_z$ along the line of R = 0.72 m in RT-1 are plotted below.

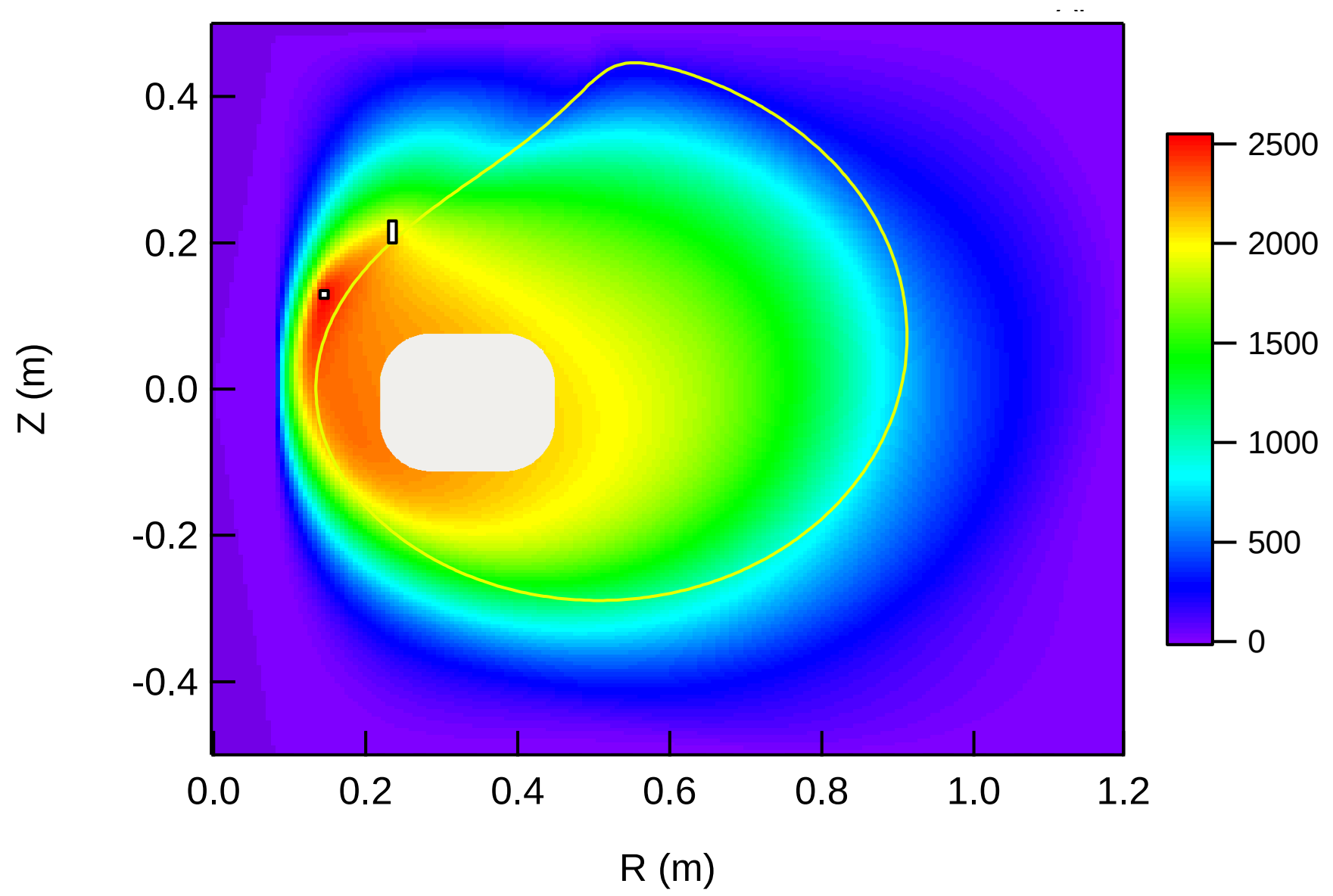

Fig. 17 Potential profile $\phi(R, Z)$ in RT-1 induced by positive charges on the double loop antenna. The $\phi(R, Z)$ is scaled to the effective voltage of the double loop antenna measured at the double loop antenna.